\documentclass[10pt]{article}
\usepackage{amsfonts,bm,bbm,amssymb,caption}
\usepackage{upgreek} 
\usepackage{mathrsfs}
\usepackage{amsmath}	
\usepackage{cite}
\usepackage{graphicx}
\usepackage{rotating}
\usepackage{hyperref}
\usepackage{verbatim}
\usepackage[all]{xy}

\csname @addtoreset\endcsname{equation}{section}
\newcommand{\beq}{\begin{equation}}
\newcommand{\eeq}{\end{equation}}
\newcommand{\bea}{\begin{eqnarray}}
\newcommand{\eea}{\end{eqnarray}}
\newcommand{\ba}{\begin{array}}
\newcommand{\ea}{\end{array}}
\newcommand{\bit}{\begin{itemize}}
\newcommand{\eit}{\end{itemize}}
\newcommand{\nn}{\nonumber}

\newcommand{\complesso}{{\ \hbox{{\rm I}\kern-.6em\hbox{\bf C}}}}
\newcommand{\reale}{{\hbox{{\rm I}\kern-.2em\hbox{\rm R}}}}

\newcommand{\p}{\partial}

\renewcommand{\a}{\alpha}

\newcommand{\g}{\gamma}

\newcommand{\G}{\Gamma}
\renewcommand{\d}{\delta}
\newcommand{\D}{\Delta}
\newcommand{\e}{\epsilon}
\newcommand{\Er}{{\mathcal{E}}}

\renewcommand{\L}{\Lambda}

\newcommand{\m}{\mu}

\newcommand{\n}{\nu}
\renewcommand{\r}{\rho}
\newcommand{\s}{\sigma}

\renewcommand{\S}{\Sigma}

\renewcommand{\t}{\theta}

\newcommand{\x}{\xi}

\newcommand{\om}{\omega}

\begin{document}

\begin{titlepage}
\begin{flushright}
CECS-PHY-13/10
\end{flushright}
\vspace{2.5cm}
\begin{center}
\renewcommand{\thefootnote}{\fnsymbol{footnote}}
{\huge \bf Pair Creation of Rotating Black Holes}
\vskip 35mm
{\large {Marco Astorino\footnote{marco.astorino@gmail.com}}}\\
\renewcommand{\thefootnote}{\arabic{footnote}}
\setcounter{footnote}{0}
\vskip 10mm
{\small \textit{
Centro de Estudios Cient\'{\i}ficos (CECs), Valdivia,\\ 
Chile\\}
}
\end{center}
\vspace{7 cm}
\begin{center}
{\bf Abstract}
\end{center}
{An exact and regular solution, describing a couple of charged and spinning black holes, is generated in an external electromagnetic field, via Ernst technique,  in Einstein-Maxwell gravity. \\
A wormhole instantonic solution interpolating between the two black holes is constructed to discuss, at the semi-classical level,  the quantum  process of creation rate, in an external magnetic field, of this charged and spinning black hole pair. }
\end{titlepage}


\section{Introduction}

In general relativity few processes are known to allow black holes creation.  At the macroscopical classical level it is possible to produce, through the gravitational collapse of stars, a massive black hole, whose mass cannot be smaller that the Oppenheimer threshold of about 3 solar masses. While at the microscopical level a quantum effect analogous to the Schwinger pair-creation in an external field may occur. That is the possibility that a spacetime with a source of excess energy will quantum tunnel into a spacetime containing two black holes.  Even though a well-established theory of quantum gravity is presently not known, some speculations about this Planckian scale effect are studied in the literature \cite{strom},  \cite{hawking}, using the semi-classical Euclidean path integral approach (not only for Einstein-Maxwell gravity but also for the dilatonic coupling \cite{dowker}, \cite{kastor}). Motivations to study this process lies, just to mention some, in the topological changing process, the black hole information paradox, black hole microstates counting and microstates interaction\cite{malda-suss}.\\
In this framework several studies were done using as a background the cosmological constant or an external magnetic field, which provide the energy to generate the black hole pair\footnote{In \cite{lemos}  is also studied the possibility of furnishing the energy to produce the black hole pair by a cosmic string in a de Sitter background.}. Of course the fact that the cosmological constant value is fixed and small  by observation, while external magnetic filed can be set arbitrary large, and actually it has recently been measured to be extremely large  at the center of some galaxies, as our own \cite{nature},  makes this process physically more realistic in the external field setting. In the first case the  Plebanski-Demianski solutions are used (see \cite{mann-ross},\cite{mann-rot}), while in the latter case the Ernst solution \cite{ernst-remove} is needed. It describes two oppositely charged black holes accelerating apart by means of  the force supplied by the external magnetic field. Ernst metrics are built by the help of generating solution  techniques. \\
Solution generating techniques (we will focus on the Ernst method \cite{ernst1}, \cite{ernst2}) are a very powerful tool in general relativity because, exploiting the integrability property of the theory, they give us a new insight into the theory and are able to generate new exact solutions, hardly obtainable by directly integrating the field equations.\\
Usually, as firstly pointed out by Gibbons in \cite{gibbons-ernst} and further analysed in \cite{strom} and \cite{hawking}, for the pair creation process, the suitable Ernst metric is the one describing a couple of accelerating, intrinsically magnetically charged black holes embedded in the external field of the Melvin magnetic universe.  The analogy with the Schwinger electron-positron pair creation in an external electric field is apparent, as discussed in section \ref{instanton}. Taking advantage of the electromagnetic duality in four-dimensions a specular treatment can also be done for an electric Reissner-Nordstrom in an external electric field. In  \cite{brown-dual} and \cite{hawking-ross}  it is shown that the pair nucleation rate of the dualised and standard cases are the same; in \cite{claudio} this electromagnetic equivalence is shown in a general setting. What is still not known is what happens in the case when the Ernst black holes are both electrically and magnetically intrinsically charged at the same time. Actually a dyonic Ernst-like solution is not even know in this case. What one expects is that the black holes acquire rotation because of  the Lorentz force interacting between the  black hole electrical charge and the the external magnetic field, as happens in the non-accelerating single black hole case \cite{ernst-magnetic}. \\
The purpose of this work is to explore the possibility of generalising the Ernst metric to the dyonic case.  This can be done taking advantage  of the new form the C-metric  offered by \cite{hong-teo}, which is more suitable for generating techniques because, in this coordinate set, the accelerating space-time can be more easily cast into the Weyl form. This is done in section \ref{generating}.  Then  in section \ref{instanton}, to evaluate the pair creation rate, an Euclidean instanton is built.

\section{Embedding  an accelerating Reissner-Nordstrom black hole in a magnetic universe}
\label{generating}

Consider Einstein gravity coupled to Maxwell electromagnetism. The regularised action for this theory is given by 
\beq \label{action}
             I[g_{\m\n}, A_\m] =  -\frac{1}{16 \pi G}  \int_\mathcal{M} d^4x  \sqrt{-g} \left( \textrm{R} - \frac{G}{\m_0}F_{\m\n} F^{\m\n} \right) - {1 \over 8\pi G}   \int_\mathcal{\p M} d^3 x \sqrt{h} \ K  - {1\over 4 \pi \mu_0} \int_\mathcal{\p M} d^3 \sqrt{h} \ n_\m A_\n F^{\m\n}  \ \  ,
\eeq
where $h$ is the determinant of the induced three metric and $K$ is the trace of the extrinsic curvature of the boundary. The first boundary term is the standard Gibbons-Hawking regularisation \cite{gibbons-hawking}, while the second is needed, for the class of solutions we will discuss,  to ensure that the electric charge is fixed on the boundary\footnote{The magnetic charge is automatically fixed by fixing the gauge potential.},  as explained in \cite{hawking-ross}.   \\
 The gravitational and electromagnetic  field equations are obtained extremising with respect to the metric $g_{\m\n}$ and the electromagnetic potential $A_\m$ respectively\footnote{Henceforward  the Newton constant $G$ and the electromagnetic vacuum permeability $\m_0$ will be set to 1 for simplicity, without loss of generality.} 
\bea  \label{field-eq}
                        &&   \textrm{R}_{\m\n} -   \frac{\textrm{R}}{2}  g_{\m\n} = \frac{2G}{\m_0} \left( F_{\m\r}F_\n^{\ \r} - \frac{1}{4} g_{\m\n} F_{\r\s} F^{\r\s} \right)  \quad ,       \\
                        &&   \partial_\m ( \sqrt{-g} F^{\m\n}) = 0   \quad .
\eea
A very well known solution for this theory is given by the dyonic Reissner-Nordstrom (RN) spacetime. It represent a static and spherical symmetric black hole equipped with both electric and magnetic intrinsic monopole charges, respectively denoted $e$ and $g$. A generalisation of RN space-time, including an acceleration parameter $A$, is called a (dyonic) charged C-metric. In spherical coordinates this metric is
\beq \label{dyon-c}
      ds^2 =  \frac{1}{(1+Ar\cos\t)^2} \left[ - Q(r) dt^2 + \frac{dr^2}{Q(r)} + \frac{r^2 d \t^2}{P(\t)} + r^2 P(\t) \sin^2\t d\varphi^2 \right] \ , 
\eeq
where
\bea \label{Qr}
         Q(r) &=& (1-A^2 r^2) \left( 1 - \frac{2m}{r} + \frac{e^2+g^2}{r^2} \right)  \quad  ,   \\ 
          P(\t) &=& 1 + 2 m A \cos \t + A^2 \cos^2\t (e^2+g^2) \quad . \label{Pt}
\eea
It is supported by the electromagnetic potential
\beq
           \mathcal{A} =  -\frac{e}{r} dt + g \cos \t d\varphi \ \ .
\eeq
This is usually interpreted as a couple of twins RN black holes\footnote{For the sake of generality we will always consider in this paper the dyonic charged case.} accelerating apart under the force of a string (or a strut), mathematically represented by an axial conical singularity typical of this kind of metrics, that will analysed after the magnetisation process. The metric (\ref{dyon-c})-(\ref{Pt}) will constitute the ``seed" solution of our construction. Apart the usual RN inner $r_-$ and outer $r_+$ event horizons,  (\ref{dyon-c}) has an accelerating horizon $r_A$ located at
\beq
      r_A=\frac{1}{A}      \quad \qquad , \quad \qquad r_\pm = m \pm \sqrt{m^2 -e^2 -g^2} \ \ \ .
\eeq
In order for the roots of the polynomial $Q(r)$ in (\ref{Qr}) to be ordered according with the C-metric interpretation \cite{hong-teo}, the physical parameters $m,e,g,A$ must satisfy the following relation:
 $$ 0 \le A r_- \le A r_+ \le 1 \ \ . $$  
We recall that in C-metrics, the azimuthal coordinate range has a hidden parameter C, which can be used, as in \cite{interpret-c}, to remove one of the characteristic conical singularities: $\varphi \in (-C\pi, C\pi] $.   \\
All axisymmetric space-times in Einstein-Maxwell gravity, thanks to the system integrability,  have the remarkable property of being generated, in principle, by the group of transformations $SU(2,1)$; for details see \cite{embed-hair}. One element of this group, called Harrison-Elhers transformation, is able to embed a generic spacetime in an external magnetic field \cite{ernst-magnetic}.  It can be written in this way\footnote{Hat stands for the transformed quantities.}
\beq \label{harrison-tr}
         {\Er} \longrightarrow \hat{\Er} = \frac{\Er}{1+B\mathbf{\Phi}-\frac{B^2}{4}\Er} \qquad  , \qquad   \mathbf{\Phi} \longrightarrow  \hat{\mathbf{\Phi}} = \frac{\mathbf{\Phi}+\frac{B}{2}\Er}{1+B\mathbf{\Phi}-   \frac{B^2}{4}\Er}  \qquad .
\eeq
$\Er$ and $\mathbf{\Phi}$  are the Ernst complex gravitational and electromagnetic potentials, for magnetising purposes, they are defined as
 \beq \label{def-Phi-Er} 
       \Er := f - |\mathbf{\Phi} \mathbf{\Phi}^*| + i h  \qquad \qquad ,  \qquad \qquad      \mathbf{\Phi} := A_\varphi + i \tilde{A}_t  \qquad , 
\eeq
where 
\bea
    \label{Atilde} \overrightarrow{\nabla} \tilde{A}_t &:=& - \frac{f}{\r} \overrightarrow{e}_\varphi \times (\overrightarrow{\nabla} A_t + \omega  \overrightarrow{\nabla} A_\varphi ) \qquad , \\
    \label{h-tr}    \overrightarrow{\nabla} h &:=& - \frac{f^2}{\r} \overrightarrow{e}_\varphi \times \overrightarrow{\nabla} \omega - 2 \ \textrm{Im} (\mathbf{\Phi}^*\overrightarrow{\nabla} \mathbf{\Phi} )  \qquad .
\eea
Since we are interested in axisymmetric space-times the functions $f,\omega,\g, A_t,A_\varphi$ depend only on the coordinates $(r,\theta)$. These functions for the seed solution can be obtained comparing (\ref{dyon-c}) with the most general axisymmetric metric, the Weyl-Lewis-Papapetrou one
 \beq \label{axis-metric-2-wicked}
                         ds^2 =  - f \left( d\phi - \om d t \right)^2 + f^{-1} \left[ \rho^2 d t^2 - e^{2\gamma}  \left( d \rho^2 + dz^2 \right) \right]  \qquad  :
\eeq
\bea
                      f(r,\t) = - \frac{r^2 P(\t) \sin^2 \t}{(1+Ar\cos\t)^2}   \qquad  &, & \qquad  \omega(r,\t)=0  \\ 
                      \rho(r,\t) = \frac{r \sin \t \sqrt{Q(r) P(\t)}}{(1+Ar\cos\t)^2}   \ \ \quad &, & \qquad z(r,\t) = \frac{(Ar\cos \t) \left[r+m(Ar\cos \t -1) -A(e^2+g^2)\cos \t \right]}{(1+Ar\cos\t)^2}   \nn
\eea
The differential operators can be taken as follows \footnote{The orthonormal frame is defined by the ordered triad $(\overrightarrow{e}_r,\overrightarrow{e}_\varphi,\overrightarrow{e}_\theta)$.} 
\beq
          \overrightarrow{\nabla} g(r,\theta) \propto  \overrightarrow{e}_r \sqrt{Q(r)} \p_r g(r,\t)  + \overrightarrow{e}_\theta \sqrt{P(\t)} \ \p_\theta g(r,\t) \quad .
\eeq
Then from (\ref{Atilde}) we can obtain the value of $\tilde{A}_t= e \cos \t$, therefore the seed Ernst potentials are 
\beq
         \mathbf{\Phi} = ( g + i e ) \cos \t   \qquad \quad , \quad  \qquad  \Er = - \frac{r^2 P(\t) \sin^2 \t}{(1+Ar\cos\t)^2}  - ( g^2 +  e^2 ) \cos^2 \t  \quad.
\eeq
Now we are able to apply the Harrison transformation (\ref{harrison-tr}) to get the complex potentials for the magnetised spacetime:
\bea
         \hat{\Er} &=&   \frac{- \displaystyle  \frac{r^2 P(\t) \sin^2 \t}{(1+Ar\cos\t)^2}  - ( g^2 +  e^2 ) \cos^2 \t}{\L(r,\t)}  \ \ \ ,  \\
          \hat{\mathbf{\Phi}} &=&   \frac{ (g - i e) \cos \t - \displaystyle \frac{B}{2} \left[ \frac{r^2 P(\t) \sin^2 \t}{(1+Ar\cos\t)^2}  + ( g^2 +  e^2 ) \cos^2 \t \right] }{\L(r,\t)} \ \ ,
\eea
 where
\beq
          \L(r,\t) = 1-B(g+i e) \cos \t + \frac{B^2}{4}  \left[ \frac{r^2 P(\t) \sin^2 \t}{(1+Ar\cos\t)^2}  + ( g^2 +  e^2 ) \cos^2 \t \right]   \ \ .
\eeq
Finally we have to return to the metric notation. From (\ref{harrison-tr}) is possible to find how $f$ change under the Harrison transformation:
\beq
            f(r,\t) \longrightarrow \hat{f}(r,\t) = \frac{f(r,\t)}{|\Lambda(r,\t)|^2}  \ \ .
\eeq
While from (\ref{h-tr}) we can obtain a relation to get the magnetised $\omega (r,\t)$:
\beq
            \overrightarrow{\nabla} \hat{\omega} (r,\t) = |\L(r,\t)|^2 \overrightarrow{\nabla} \omega - i \overrightarrow{e}_\varphi \times  \frac{\r}{f} ( \L^*\overrightarrow{\nabla} \L - \L \overrightarrow{\nabla} \L^* )  \ \ .
\eeq
Integrating this latter one finds that, 
\beq
         \hat{\omega}(r,\t) = \frac{e B^3 (1+2Ar\cos \t) Q(r)}{2 A^2 r (1+Ar \cos \t)^2} + \frac{eB}{2A^2r} \left[ 4A^2 + B^2A^2(e^2+g^2) -B^2\right] + \frac{eB^3m}{A^2 r^2} - \frac{e B^3(e^2+g^2)}{2A^2r^3} +\om_0 \
\eeq
where $\omega_0$ is an arbitrary constant. From definition (\ref{def-Phi-Er}) we have
\bea
   \label{a_phi_hat}      \hat{A}_\varphi (r,\t) &=&   \frac{g\cos \t - B (e^2+g^2) \cos^2 \t -\displaystyle \frac{B}{2} \left(\frac{3 g B}{2} \cos \t -1 \right) \Er  -\frac{B^3}{8}  \Er^2 } {|\L(r,\t)|^2}  +  k_\varphi \quad ,
          \\
         \hat{\tilde{A}}_t (r,\t) &=& e \cos \t  \frac{ \left\lbrace 1- \displaystyle \frac{B^2}{4}  \left[ \frac{r^2 P(\t) \sin^2 \t}{(1+Ar\cos\t)^2}  - ( g^2 +  e^2 ) \cos^2 \t \right] \right\rbrace }{|\L(r,\t)|^2}   +\tilde{k}_t   \quad ,
\eea
Using (\ref{Atilde}) it is possible to obtain the standard electric field component:
\beq \label{a_t_hat}
          \hat{A}_t(r,\t) = -\hat{\omega}(r,\t) \left[ \hat{A}_\varphi(r,\t) + \frac{3}{2B}  \right] + \frac{2e}{r}  + k_t \quad ,
\eeq
where $k_t$, $\tilde{k}_t $ and $k_\varphi$ are generic integration constants.\\
Finally inserting the Harrison transformed quantities ($\hat{f}$ and $\hat{\omega}$, while $\g$ remains unvaried) in (\ref{axis-metric-2-wicked}), the C-metric solution (\ref{dyon-c}) magnetised, supported by the electromagnetic field (\ref{a_t_hat})-(\ref{a_phi_hat}), results:
\beq \label{magn-dyon-c}
      d\hat{s}^2 =  \frac{|\L(r,\t)|^2}{(1+Ar\cos\t)^2} \left[ - Q(r) dt^2 + \frac{dr^2}{Q(r)} + \frac{r^2 d \t^2}{P(\t)} \right] + \frac{r^2 P(\t) \sin^2\t \left[d\varphi- \omega (r,\theta) dt \right]^2}{|\L(r,\t)|^2 (1+Ar\cos\t)^2}  \ .
\eeq
This metric describes a pair of spinning, Reissner-Nordstrom dyonically and oppositely charged black holes accelerating away from each other along the axis of a  magnetic universe. Remarkably, even though the seed solution was diagonal, (\ref{magn-dyon-c}) exhibits rotation due to the appearance of a $\overrightarrow{E} \times \overrightarrow{B}$ circulating momentum flux in the stress-energy tensor, which serves as a source for a twist potential. This is a typical feature of magnetised black holes when the spacetime possesses intrinsic charge and external electromagnetic field of different type (i.e. electric intrinsic charge and external magnetic field, or vice-versa), see for instance \cite{ernst-magnetic}. That's because the Ernst potentials are fully complex (not just real or purely imaginary). In fact the metric  (\ref{magn-dyon-c}) is the rotating generalisation of the one found by Ernst in \cite{ernst-remove} and studied in \cite{strom} - \cite{hawking}. This latter sub-case can be obtained from (\ref{magn-dyon-c}) by setting $e=0$, that is retaining only the intrinsic magnetic charged black hole. This is why the $e=0$ case has no rotation.\\
Due to the accelerating and magnetised asymptotic,  it is not known how to compute the angular momentum for these magnetised spacetimes. In case of no acceleration, so just for a single black hole, a couple of recent results are known, but they disagree\footnote{While the work \cite{gibbons-pope2} refers exactly to the theory we are treating in this paper, i.e.  Einstein-Maxwell gravity, \cite{yaza} considers a slightly different coupling involving also a scalar dilaton. This could be the reason of the discordance.}. As commented in \cite{yaza} the gravitational contribution to the angular momentum is exactly compensated by the contra-rotation of the external electromagnetic field. Therefore, even though the charged black hole in an external magnetic field is rotating, the total angular momentum of the space-time is null. While as computed in \cite{gibbons-pope2}, the angular momentum is not vanishing.  \\
The spacetime (\ref{magn-dyon-c}), as usually occurs for accelerating metrics, is affected by conical singularities, which actually act as the sources of the acceleration\footnote{Also the seed metric (\ref{dyon-c}) has axial deficit/excess angles, which can be quantified in the following computation just turning off the external magnetic field: $B=0$. }. To study the metric conicity, following \cite{interpret-c}, a small circle around the half-axis $\t=0$  is considered (while keeping the coordinates $t$ and $r$ fixed):
\beq \label{cir/rad}
{\hbox{circumference}\over\hbox{radius}} 
 =\lim_{\t \to 0} {2\pi C P(\t) \sin \t \over \t \ |\L(r,\t)|^2} = \frac{ 2 \pi C \left[1+ 2mA + A^2 (e^2+g^2) \right]}{e^2 B^2+\left[ 1- gB + \frac{B^2}{4} (e^2+g^2) \right]} \ \ .
\eeq
To avoid the conical singularity in $\t=0$ the parameter $C$ can be fixed such that
\beq \label{c-const}
           C =  \frac{ e^2 B^2+\left[ 1- gB + \frac{B^2}{4} (e^2+g^2) \right]}{1+ 2mA + A^2 (e^2+g^2) } \ \ .
 \eeq 
Then the coupling between the intrinsic charges and the external magnetic field allows us to regularise the nodal singularity around $\t=\pi$. In fact,  imposing the lack of deficit or excess angle at $\t=\pi$, as done in (\ref{cir/rad}), we get a constraint relation between the physical parameters $e,g,m,A$ and $B$:
\beq \label{reg-const}
       \frac{\big[ 1 + gB + \frac{B^2}{4} (e^2+g^2) \big]^2 + e^2 B^2 }{\left[ 1- gB + \frac{B^2}{4} (e^2+g^2) \right]^2 + e^2 B^2} \cdot \frac{1- 2mA + A^2 (e^2+g^2) }{1+ 2mA + A^2 (e^2+g^2) } = 1 \qquad .
\eeq
This means that the force necessary to accelerate the two black holes is entirely provided by the external magnetic field, without any need for a pulling string. \\ 
In the case without the intrinsic electric charge, the non relativistic limit of this constraint, that is for small acceleration $A\approx 0$, describes the Newtonian force felt by a massive magnetic monopole, of intensity $g$, in a uniform magnetic field of strength $B$. That approximation corresponds, in fact, to the weak magnetic field limit,
\beq
          mA \approx -gB \ \ .
\eeq
 The addition of the intrinsic electric charge to the black hole leaves this limit unchanged, because $eB$ is a subleading contribution, which is only relevant at higher orders. From a Newtonian point of view this is related to the fact that the Lorentz force for an electrically charged particle in an external magnetic field is proportional to both the magnetic field and the speed of the particle, which, since in the non relativistic limit the speed is small, produce a further factor of damping. \\
Anyway, even away from the non relativistic limit,  note how the intrinsic magnetic field plays a prominent role in the angular regularisation, because when $g=0$ in (\ref{reg-const}) both conical singularities can be eliminated only in the particular cases of vanishing mass parameter ($m=0$), or trivially in the vanishing acceleration case ($A=0$). Of course the role of intrinsic electromagnetic charges can be switched, without changing the form of the metric, by an electromagnetic duality rotation, that is embedding the dyonic black hole in an external electric field.  \\
Therefore the spacetime (\ref{magn-dyon-c}) implemented by the constraint (\ref{reg-const}) is completely regular, since the only remaining singularities, of curvature, are located inside the inner horizon at $r=0$. \\
At spatial infinity, that is for $\t \rightarrow \pi, r \rightarrow A^{-1}$, the solution is not asymptotically Melvin as the $e=0$ case. This is a typical feature of spinning magnetised black holes \cite{hiscock}, \cite{gibbons-pope}.  To see it, just consider that the value of the electric field is not converging to zero as in the Melvin universe. \\
In case of null acceleration the metric describes just a single spinning RN black hole embedded in an external magnetic field \cite{ernst-magnetic}\footnote{For a notation similar to the one used here see also appendix A in \cite{marcoac}, fixing $s=0$.}.

\section{Instantonic pair creation}
\label{instanton}

In order to interpret the solution as a black hole pair creation, and evaluate its nucleation rate, as discussed in  \cite{hawking},  \cite{dowker} and  \cite{kastor} the $(y,x)$ coordinate are always preferred to the spherical ones. They are related by the following transformation
$$  T= At \ ,\qquad  y = - \frac{1}{Ar}  \ , \qquad x = \cos \t   \ \ .$$

 In this set of coordinates  the dyonic magnetised C-metric solution (\ref{magn-dyon-c}) becomes\footnote{A Mathematica notebook checking this metric can be found at \url{https://sites.google.com/site/marcoastorino/papers/1312-1723}. }
\beq    \label{magn-dyon-c-yx}
            ds^2= \frac{|\L(y,x)|^2}{A^2 (x-y)^2} \left[ G(y) d T^2 - \frac{dy^2}{G(y)} + \frac{d x^2}{G(x)} \right] + \frac{ G(x) \left[d\varphi- \omega (y, x) d T \right]^2}{|\L(y,x)|^2 A^2 (x-y)^2}  \ ,
\eeq
where
\bea 
G(\xi) &=&   (1-\xi^2)(1+r_-A\xi)(1+r_+A\xi) \\
\L(y,x) &=& 1 -B x (g-i e) + \frac{B^2}{4} \left[ \frac{G(x)}{A^2(x-y)^2} + (e^2+g^2)x^2 \right]  \\
\omega(y,x) &=& \frac{B^3e(y-2x)}{2 A^2 (x-y)^2} G(y) - \frac{Bey}{2A^2} \left[ 4A^2 - B^2 + A^2B^2(e^2+g^2)\right] +  \frac{B^3 e m y^2}{A} + \frac{B^3 e y^3}{2}  (e^2+g^2) +\omega_0 \nn  \\
A_\varphi(y,x) &=& \frac{ \left\{gx -  \frac{B}{2} \left[ \frac{G(x)}{A^2(x-y)^2} + (e^2+g^2)x^2 \right] \right\} \left\{1 -  gxB + \frac{B^2}{4} \left[ \frac{G(x)}{A^2(x-y)^2} + (e^2+g^2)x^2 \right] \right\} -Be^2x^2  }{|\L(y,x)|^2} +k_\varphi\ \nn \\
A_T(y,x) &=&  - \omega(y,x) \left[A_\varphi(y,x) + \frac{3}{2B}\right] - 2 e y +k_T \label{AT}
\eea 
In this new set of coordinates the non accelerating limit is not as explicit as in  (\ref{magn-dyon-c}) and also the geometrical interpretation is clearer in spherical coordinates. On the other hand it is clear that (\ref{magn-dyon-c-yx})-(\ref{AT}) is the generalisation of the metric considered in \cite{strom} and \cite{hawking}, which can be obtained vanishing the electric charge $e$. There is only  a subtle difference in the parametrization of the polynomial $G(\xi)=-r_-r_+A^2\prod_{i=1}^4 (\xi-\xi_i)$, according to the insight of \cite{hong-teo}, therefore the roots don't always coincide, it means that the location of the horizons and range of the coordinates may differ.
The angular coordinates are $(x,\varphi)$ and in order for the metric to have Lorentz signature we require $\xi_3 \le x \le \xi_4$, so that the sign of $G(x)$ is positive. Because of the conformal factor $1/(x-y)^2$ in the metric  the spatial (and conformal) infinity is reached by fixing $t$ and letting both $y$ and $x$ approach $\x_3$. The inner, event and accelerating horizons are located at $y=\x_1$, $y=\x_2$ and $y=\x_3$ respectively. While the $x=\x_3$ axis is pointing towards spatial infinity, the $x=\x_4$ one is pointing towards the other black hole. 
Usually the constant $k_\varphi$ is fixed in order to confine the Dirac string of the magnetic field to the axis $x=\xi_4$. This can be accomplished by fixing $k_\varphi$ so that $A_\varphi(x=\x_3)=0$. \\
Similarly to the case for $e=0$ \cite{hawking} to ensure that the metric is free from conical singularities, on both poles $x=\x_3,\x_4$ we have to impose  

\beq \label{nocone3}
          G'(\xi_3)|\L(\xi_4)|^2=-G'(\xi_4)|\L(\xi_3)|^2
\eeq
and
\beq \label{nocone4}
          \D \varphi_E = \frac{4\pi |\L(\xi_3)|^2}{G'(\xi_3)} \ \ ,
\eeq

which are precisely equivalent to the constrains  (\ref{c-const}) and  (\ref{reg-const}), we have previously obtained in spherical coordinates. Note that $\L(\xi_i):=\L(x=\xi_i)$ are just constants. \\

The black holes pair production probability $|\Psi|^2$ is described\footnote{Up to a normalization factor.}, according to the no-boundary Hartle and Hawking proposal \cite{hartle}, by the functional integral over all possible manifold topologies, metrics and electromagnetic potentials interpolating between two boundary hypersurfaces $\S_1$, $\S_2$. The amplitude for this process is given by the wave function
\beq
          \Psi_{12} = \int  \mathscr{D} [\mathcal{M}]  \mathscr{D} [g] \mathscr{D} [\mathcal{A}] \ \exp(-i \ I [\mathcal{M} , g, \mathcal{A}])  \ \ .
\eeq
The measure $ \mathscr{D} [\mathcal{M}]  \mathscr{D} [g] \mathscr{D} [\mathcal{A}]$ on the functional space is not well defined and however, even if properly defined, it would be computationally impractical to handle. Fortunately, in analogy with the flat case, we can make use of a semi-classical simplifying assumption which relies on the existence of an instanton. An instanton is an Euclidean regular solution which interpolates between the initial $(1)$ and final $(2)$ states of a classically forbidden transition. It is a saddle point for the Euclidean path integral that describes the pair nucleation probability, so the transition probability amplitude is well  approximated, at the lowest order in the Planck length, by
\beq
             \Psi_{12} \approx e^{-I_e}  \ \ .
\eeq 
 Therefore, to obtain the pair creation rate between the two black holes described by the C-metric and their magnetic background, one has to build the instanton from  (\ref{magn-dyon-c-yx}).  $I_e$ is a real action evaluated on a Riemannian solution of the Einstein-Maxwell equations (\ref{field-eq}), which does not have necessarily  to be real. For a rotating solution such as ours, we can choose whether consider a real or a complex instanton.  It is argued in \cite{mann-rot}, \cite{mann-booth}, \cite{brown-dual} that for this stationary pair production the latter is more physical, because to enforce reality one has to impose some charge parameters to be imaginary. It means that, in that case, the Euclidean and Lorentzian solutions do not properly match, because the positions of the horizons are different, or even worst some horizons may even disappear because the number of real roots may change. Moreover neither electromagnetic charge nor angular momentum would be conserved in the pair production. But, maybe, the more problematic point consists in the fact that the extrinsic curvature, induced metric and the induced electromagnetic field will not match on the spatial hypersurface joining the Euclidean and the Lorentzian solutions. Thus  the introduction of extra thin wall matter would be necessary  to fix this issue.  For these reasons we will consider the possibility of having a complex instanton, provided that the action evaluated on this solution is real, thus the creation probability remains real as well.\\
 A standard way (as done in \cite{kastor}, \cite{hawking}, \cite{dowker}) to generate the instanton is to Euclideanise the Ernst solution (\ref{magn-dyon-c-yx}) by setting $\tau=iT$ and then fix the Euclidean period to regularise the conical singularity in the  ($y,\tau$) section. In \cite{mann-rot} (see also \cite{brown-dual}), it is shown that this is equivalent to requiring regularity to the extrinsic curvature $K_{ij}$, the induced metric $h_{ij}$ and the induced electromagnetic field $(E_i,B_i)$ on the gluing space-like hyper surfaces $\S_\tau$, defined by constant $\tau$. In this $3+1$ foliation the Euclidean spacetime takes the usual form
\beq
           ds^2= N^2 d\tau^2 + h_{ij} (dx^i+iV^i d\tau) (dx^j+iV^j d\tau)     \quad , 
\eeq 
where $N$ and $V^j$ are the lapse function and the shift vectors, which can depend only on $(x,y)$ coordinates to respect axisymmetry. The prescription for the supporting electromagnetic field is given by
\beq
         F_{tj} = i \tilde{F}_{tj} \ , \qquad F_{jt} = i \tilde{F}_{jt} \ , \qquad  F_{jk} =  \tilde{F}_{jk} \quad .
\eeq 
Since the metric is complex, its signature is not clearly defined, so we can adopt the meaning of Euclidean from  \cite{mann-rot}, if at any point $x_0^\a$ there exists a complex spatial-coordinate transformation $x^j = \tilde{x}^j -i V^j(x_0^\a) \tau$ that, absorbing the shift vector $V^j$, puts the metric in Euclidean diagonal form  
\beq
          ds^2\Big|_{x=x_0} = N^2 d\tau^2 + h_{ij}d\tilde{x}^id\tilde{x}^j \qquad .
\eeq
We are interested in the lukewarm solution, which is defined as having the event and accelerating horizons not degenerate and at thermal equilibrium. Therefore we have to impose that the surface gravity and the temperature are the same on both horizons $y=\x_2$ and $y=\x_3$. This can be implemented by constraining further the structure constants of the black hole to avoid the conical singularity on the $(y,\tau)$ section. This is done in two steps: first fixing the period of the Euclidean time to be
\beq \label{reg-xi3}
          \D\tau= {4\pi \over G'(\xi_3)} \quad ,
\eeq 
on the accelerating horizon $y=\xi_3$, and then requiring that this value coincides with the one of the event horizon in $y=\xi_2$, that is
\beq \label{reg-xi4}
                   G'(\xi_2) = - G'(\xi_3) \quad .
\eeq 
These conditions on $G(\xi)$ are formally identical to the electrically neutral case (i.e. non rotating) studied in \cite{hawking}. The basic difference is that the $G(\xi)$ differs respect to the $e=0$ case, mainly in the horizon positions. Moreover we have one more parameter (the one related with the intrinsic electric charge) to accomplish the regularity constraints (\ref{nocone3}), (\ref{nocone4}), (\ref{reg-xi3}), (\ref{reg-xi4}). For this reason, in the pair creation process, more general black hole can be produced with respect to the static case \cite{hawking}.  \\
From (\ref{reg-xi4}) one gets
\beq
          (\xi_4 - \xi_3)(\xi_3 - \xi_1 ) = (\xi_4-\xi_2) (\xi_2-\xi_1)   \quad ,
\eeq 
which, in the non degenerate case $\xi_3\ne \xi_2$, can be further simplified to
\beq \label{con-costr}
           \xi_4 - \xi_3 = \xi_2-\xi_1   \ \ \ .
\eeq 
In terms of the physical parameters equation (\ref{con-costr}) means 
\beq
          A(e^2+g^2) = \sqrt{m^2-e^2-g^2}  \ \ .
 \eeq  
This condition further restricts the regularity constraint (\ref{reg-const}), and hence also the pair production process. \\
The resulting instanton has topology $S^2\times S^2-\{pt\}$, where the removed point is $y=x=\xi_3$. In the literature this instanton is interpreted as representing the creation, in an external magnetic field, of a pair of oppositely charged black holes which subsequently uniformly accelerate away from each other \cite{hawking}, \cite{kastor}. The two black holes are  connected by a  wormhole throat, containing the event horizon, which is located at finite proper distance from the wormhole mouth.\\
To compute the black hole pair creation rate we need to evaluate the action (\ref{action}) on the instanton we have just built and compare with the value of the action on the background.  
Using the trace of the equation of motion  (\ref{field-eq}) and using the Stokes theorem for the Maxwell term $F^2$, the  action (\ref{action}) can be recast, on shell, as a boundary term
\beq \label{boundary-term}
          I = - {1 \over 8 \pi G} \int_{\p \mathcal{M}} d^3x \sqrt{h}  \ [ F^{\m\n} n_\m A_\n+ \nabla_\m n^\m  ]  \ \ ,
 \eeq
where $n^\m$ is the normalised vector orthogonal to the boundary surface $y=x=\xi_3$ . Explicitly one evaluates the action at $y=x-\e$ and then takes the limit $\e \rightarrow 0$, so $\e$ acts as a regularisation. The non-null components of the unit outward pointing normal to the surface $y=x-\e$ are
\beq
                 n^y = - { A (x-y) G(y)  \over | \L(y,x) | \sqrt{G(x)-G(y)}  } \qquad , \qquad n^x = - { A (x-y) G(x)  \over | \L(y,x) | \sqrt{G(x)-G(y)}  }  \qquad ,
\eeq
while the instantonic induced three-metric on the  $y=x-\e$ hypersurface is
\beq
                      d\hat{s}^2= \frac{|\L(y,x)|^2}{A^2 (x-y)^2} \left[ -G(y) d \tau^2 +  \frac{G(x) -G(y)}{G(x) G(y)} dx^2 \right] + \frac{ G(x) \left[d\varphi +  i \omega (y, x) d \tau \right]^2}{|\L(y,x)|^2 A^2 (x-y)^2} \Big|_{y=x-\e} \quad .
\eeq
Hence, according to the no-boundary proposal,  the creation rate of the dyonic RN black hole pair with respect to the background (bkgr) is given by:
\beq
              \G_{dyonRN \over \text{bkgr}} \propto { |\Psi_{dyonRN} |^2\over |\Psi_\text{bkgr}|^2 } \propto \textrm{e}^{-2(I_{dyonRN}-I_\text{bkgr})}
\eeq
Unfortunately in the case of rotating Ernst metrics, the behaviour at infinity is not clear, see \cite{gibbons-pope} for recent developments, so in our case evaluating the background contribution is problematic. At most we might speculate that the naive regularisation carried out in \cite{dowker}, consisting in eliding the divergent term with the background contribution, also works in the rotating case\footnote{Even though the result of \cite{dowker} is eventually correct, some subtleties in this asymptotic regularisation process have to be carefully considered, as analysed in \cite{hawking}.}. \\
The analogy with the Schwinger electron-positron production (of charge $\pm\hat{e}$)  in an external electric field ($\hat{E}$) is manifest when the intrinsic black hole electric charge $e$ is null, therefore the RN black hole couple is not spinning (Ernst potentials and the instanton are real). In that case the boundary action (\ref{boundary-term}), according to \cite{hawking-ross}, reduces to\cite{hawking},\cite{dowker}:
\beq \label{int-delta}
          I= - {1 \over 8 \pi} \int_{\p \mathcal{M}} d^3x \sqrt{h} \  e^{-\d} \nabla_\m (e^\d n^\m)  \quad ,
\eeq 
where   $e^{-\d}=\L{(y-\xi_1) \over (x-\xi_1)}$. Then performing the trivial integrations over $\tau$ and  $\varphi$, the action evaluated on the instanton becomes
\beq \label{last-int}
            I=  - {1 \over 8 \pi} \D\tau  \Delta \varphi_E  \int_{\xi_3}^{\xi_3+\e} dx  { \sqrt{h} \over \sqrt{g} }  e^{-\d} \p_\m (e^\d \sqrt{g} n^\m)   \Big|_{y=x-\e} \quad .
\eeq 
Expanding in powers of $\e$ and integrating (\ref{last-int}) we obtain\footnote{Note that there seems to be a factor 1/2 discrepancy in $I_{Ernst}$ between the references, of the same authors \cite{hawking} and \cite{hawking-ross}, even though they claim to obtain the same result. The final result of $\G$ is obtained exactly changing the definition of the pair creation rate, from \cite{hawking} to \cite{hawking-ross}, by a compensating factor 2.}
\beq
         I = I_0 +  \frac{\L^2(\xi_3)}{A^2 G'(\xi_3)}  \frac{\pi}{\xi_3 - \xi_1}  \qquad .
\eeq
The first factor $I_0$ is divergent on the boundary, when $\e \rightarrow 0$, but evaluating the pair production rate relative to the Euclidianised  magnetic universe, this term is compensated by the background contribution. In fact this can be checked evaluating, up to the order $O(\e)$, the action \label{int-delta} on the Melvin background, setting $r_\pm=0$ on the instanton metric. Finally the production rate of a pair of not spinning, magnetically charged RN black holes, in an external magnetic field background, with respect to the Melvin background is given by
\beq \label{rate}
             \G_\frac{RN}{Melvin} \propto \exp \left[ \frac{-2 \pi \ \L^2(\xi_3)}{A^2 G'(\xi_3)(\xi_3 - \xi_1)}  \right] \quad .
\eeq
In terms of the value of the magnetic field at infinity (which coincides with the Melvin background magnetic field)  $\hat{B}=\frac{B}{2} \frac{G'(\xi_3)}{\sqrt{\Lambda(\xi_4)}}$ and the  intrinsic physical magnetic charge 
\beq
          \hat{q} := { 1 \over 4\pi} \int_\S F_{\m\n} dx^\m\wedge dx^\n = { 1\over 4\pi} \int_0^{\Delta\varphi_E} d\varphi \int_{\xi_3}^{\xi_4} dx \ \p_x A_\varphi = g \ {\xi_4-\xi_3\over G'(\xi_3)} \ \frac{\L^{3/2}(\xi_3)}{\L^{1/2}(\xi_4)} \quad ,
 \eeq 
where $\S$ is any two sphere surrounding the black hole horizon, (\ref{rate}) can be rewritten as 
\beq \label{pprod}
                \G_\frac{RN}{Melvin} \propto \exp \left[  -4  \pi \hat{q}^2  \frac{(1-\hat{B} \hat{q})^2}{1-(1-\hat{q} \hat{B})^4}  \right] \ \ .
\eeq 
Expanding (\ref{pprod}) for small $\hat{q} \hat{B}$ we obtain a similar behaviour with respect to the leading term of the Schwinger pair production $\pi m^2/\hat{e}\hat{E}$:
\beq 
            \G_\frac{RN}{Melvin}  \approx   \exp \left[ - \hat{q}^2 \left( \frac{\pi}{\hat{B} \hat{q}} - \frac{\pi }{2} + ... \right) \right] \quad.
\eeq
Therefore, even though the new C-metric parametrisation introduced in \cite{hong-teo} is not completely physically equivalent to the older one\footnote{In case of rotation this coordinate system makes the accelerating Kerr black hole free from torsion singularities,  that generates closed time-like curves \cite{hong-teo2}.}, the pair creation rate remains the same as \cite{strom}, \cite{hawking}, \cite{dowker}, at least in the no rotating case. 
\\

\section{Comments and Conclusions}

In this paper a generalisation of the Ernst metric describing a couple of accelerating, intrinsically electrically and magnetically charged black holes in the presence of an external electromagnetic field is generated, by means of the Ernst's solution generating technique. The main novelty with respect to the only intrinsically magnetically charged case, consists in the fact that the presence of the electric charge embedded in an external magnetic field makes, due to the Lorentz force, the Reissner-Nordstrom black hole pair rotate. Then, thanks to the presence of the external magnetic field, it is possible to regularise the conical singularity typical of these accelerating solutions. Therefore there is no need of a cosmic string or a strut to provide the acceleration, but the acceleration is furnished by the external magnetic field.\\
The relevance of this result lie in the fact that this is a completely regular, analytic, rotating, two-black hole exact solution. Up to our knowledge, it represents the first example of such kind in the theory of pure Einstein-Maxwell general relativity.    \\
Thereafter from this metric an instantonic solution is built. It interpolates between the two classical states consisting in the black hole pair and its magnetic background. Being a saddle point for the Euclidean path integral, it is used to describe, at the semi-classical level, the quantum nucleation probability between the two forbidden classical states. This is in close analogy with the Schwinger electron-positron pair creation in an external electric field. The instanton considered here is of more general type with respect to the usual studied in the literature because it includes also the electric charge. \\
A better understanding of the asymptotic behaviour may be useful to clarify the charges of the solution here considered and also to evaluate the contribution of the rotation on the pair creation rate.\\ 
It would be also interesting to extend this analysis starting, as a seed, with a more general black hole pair, which includes rotation from the beginning, that is an accelerating Kerr-Newman metric. It would generalise and unify, at the same time, both the Ernst's family of solutions describing black holes embedded in an external (electro)magnetic field and the Plebanski-Demianski family, which, on the other hand, describes accelerating metrics. Including six physical parameters: mass rotation, acceleration, external electro-magnetic fields, intrinsic electric and magnetic charges, it would represent the most generic, physical, black hole metric for electro-vacuum general relativity\footnote{We mean Maxwell electromagnetism coupled with pure general relativity without cosmological constant, because in presence of cosmological constant some symmetries, such as the Harrison transformation, are broken\cite{ernst-lambda}. }. Works in this direction are in progress.

\section*{Acknowledgements}
\small I would like to thank Eloy Ay\'{o}n-Beato, Hideki Maeda, Cristi\'{a}n Mart\'{i}nez, Gianni Tallarita and Cedric Troessaert  for fruitful discussions. 
\small This work has been funded by the Fondecyt grant 3120236. The Centro de Estudios Cient\'{\i}ficos (CECs) is funded by the Chilean Government through the Centers of Excellence Base Financing Program of Conicyt.
\normalsize


\appendix



\end{document}